
\input harvmac
\input epsf.tex
\input tables.tex
\ifx\answ\bigans

\else

\fi

\def\caption#1{
	\centerline{\vbox{\baselineskip=12pt
	\vskip.15in\hsize=3.8in\noindent{#1}\vskip.1in }}}
%
%
\def\INT#1#2{\noindent#1\hfill #2%
\bigskip\supereject\global\hsize=\hsbody%
\footline={\hss\tenrm\folio\hss}}
\def\abstract#1{\centerline{\bf Abstract}\nobreak\medskip\nobreak\par #1}
\def\np#1#2#3{Nucl. Phys. B{#1} (#2) #3}
\def\pl#1#2#3{Phys. Lett. {#1}B (#2) #3}
\def\prl#1#2#3{Phys. Rev. Lett. {#1} (#2) #3}
\def\physrev#1#2#3{Phys. Rev. {#1} (#2) #3}

\def\tanb{\tan \beta}
\def\frac#1#2{{\textstyle{#1 \over #2}}}
\def\[{\left[}
\def\]{\right]}
\def\({\left(}
\def\){\right)}
\def\address{\vbox{\sl\centerline{Institute for Nuclear Theory}
\centerline{University of Washington, NK-12}
\centerline{Seattle, WA  98195}}}

\nref\largetanb{S. Dimopoulos, L. J. Hall and S. Raby,
\prl{68}{1992}{1984};  \physrev{D45}{1992}{4192};
\physrev{D47}{1993}{R3697}.}

\nref\attempts{J. Harvey, D. B. Reiss and P. Ramond,
\np{199}{1982}{223}; K. S. Babu and R. N. Mohapatra,
\prl{70}{1993}{2845}; E. Papageorgiu, preprint
LPTHE-ORSAY-40-94 (1994) hep-ph-9405256.}

\nref\anderson{G. Anderson {\it et al.}, \physrev{D49}{1994}{3660}.}

\nref\famsym{P. Frampton and T. Kephart, preprint IFP-702-UNC
(1994) hep-ph-9409330; M. Leurer, Y. Nir, N. Seiberg,
\np{398}{1993}{319};
P. Pouliot and N. Seiberg, \pl{318}{1993}{169};
G. Lazarides and Q. Shafi, \np{350}{1991}{179};
\np{364}{1991}{3}; Y. Achiman and T. Greiner, \pl{329}{1994}{33};
M. Joyce and N. Turok, \np{416}{1994}{389};
C. Froggatt and H. Nielsen, \np{147}{1979}{277}.}

\noblackbox
\vskip 1.in
\centerline{{\titlefont Neutrino Oscillations from Discrete}}
\vskip.2in
\centerline{\titlefont  Non-Abelian Family Symmetries }
\bigskip\bigskip\bigskip
\centerline{Martin Schmaltz}
\bigskip\address
\vfill\abstract{
I discuss a SUSY-GUT model with a non-Abelian discrete family
symmetry that explains the observed hierarchical pattern of quark
and lepton masses. This $SO(10) \times \Delta(75)$ model predicts
modified quadratic seesaw neutrino masses and mixing angles
which are interesting for three reasons: i.) they offer a solution to
the solar neutrino problem,  ii.)  the tau neutrino has the right mass
for a cosmologically interesting hot dark matter candidate, and iii.)
they suggest a positive result for the $\nu_\mu \rightarrow \nu_\tau$
oscillation searches by the CHORUS and NOMAD collaborations.
However, the model shares some problems with many other
predictive GUT models of quark and lepton masses. Well-known
and once successful mass and angle relations, such as the $SU(5)$
relation $\lambda_b^{GUT}=\lambda_\tau^{GUT}$, are found to be
in conflict with the current experimental status. Attempts to correct
these relations seem to lead to rather contrived models. }

\vfill\INT{INT94-00-74,\ hep-ph/9411383}{November 1994}

\newsec{Introduction}

Fermion masses and mixing angles correspond to free parameters
in the Standard Model (SM). It is widely believed that there should
be a more general theory that predicts at least some of these
parameters from first principles. Even though this problem has
inspired many theorists to attempt a solution
\refs{\largetanb-\famsym}\  we are still lacking a compelling theory.
The obvious hierarchical pattern of the masses and mixing angles
seems to suggest a possible explanation via a slightly broken symmetry \famsym.

It has been shown in a previous publication \ref\ours{D. Kaplan and M.
Schmaltz, \physrev{D49}{1994}{3741}.}\  that a non-Abelian family
symmetry, with the three families transforming as an irreducible
representation, can be used as a very powerful tool to constrain the
Yukawa couplings of the SM, resulting in interesting fermion mass
textures. In \ours, it has been demonstrated that these symmetries
naturally suppress flavor changing neutral currents in
supersymmetric theories. Kaplan and Muyarama have used
non-Abelian symmetries to constrain ``dangerous" proton decay
operators \ref\protondecay{D. Kaplan and H. Murayama,
\pl{336}{1994}{221}.}.

The model presented as an example in \ours\  demonstrates nicely
how interesting Yukawa matrix textures can be obtained from
non-Abelian family symmetries.
Unfortunately, due to the large number of unknown parameters
entering the Yukawa coupling matrices, it does not give rise to any
precise numerical predictions.

The model presented in this publication is based on the same
approach - it is an $SO(10)$ SUSY-GUT with a non-Abelian family
symmetry\foot{In addition, a flavor blind $U(1)$ or $R$ symmetry is
required in order to forbid some unwanted couplings.} - but is more
ambitious: it predicts the light quark masses ($m_s$, $m_d$, $m_u$)
as well as all neutrino mass ratios and lepton mixing angles. The
reduction of parameters in this model relative to the one in \ours\  is
due to a more efficient exploitation of the restrictive power of the
$SO(10)$ gauge symmetry.

The $\Delta(75)$  family symmetry of the model determines the
Yukawa matrix texture. At the GUT scale the three families are unified
into the fundamental triplet representation of $\Delta(75)$. Below
$M_{GUT}$ the family symmetry is broken and the hierarchical
pattern of Yukawa couplings is generated. The coupling strengths
are determined by the charges of the various fields under the $Z_3$
and $Z_5$ subgroups of $\Delta(75)$. These charges allow only the
top quark to have a renormalizable coupling to an $SU(2) \times
U(1)$ breaking Higgs VEV; all other couplings arise at higher orders
of $\Delta(75)$ breaking. The spontaneous family symmetry breaking
is accomplished with a few non-trivial Higgs VEVs.

Once created by the family symmetry at the GUT scale, the
hierarchical coupling patterns are protected by the
non-renormalization property of supersymmetry.

The most interesting predictions of this model lie in the neutrino
sector. The model, which has been constructed to fit the SM fermion
masses and mixing angles, has a completely determined neutrino
Dirac mass matrix $Y_\nu$. All its components are related to entries
in the quark and charged lepton matrices by the $SO(10)$ symmetry.
Since the non-Abelian family symmetry constrains the Majorana
mass matrices for the right handed neutrinos to be very simple (in
this model, it is proportional to the unit matrix) one can
unambiguously predict all the neutrino mass ratios and lepton
mixing angles via the seesaw approximation \ref\seesaw{M.
Gell-Mann, P. Ramond, R. Slansky  in {\it Supergravity} (ed.
D.Z. Freedman and P. van Nieuwenhuizen)
North Holland, Amsterdam.}. The  $SO(10)$ Clebsches modify the
usual quadratic mass relations in an interesting way.
One finds that
\item{i.)} the predictions for $\sin^2(\Theta_{\nu_e\mu}) = 0.019 \pm
0.008$ and $m_{\nu_\mu} \sim \CO(10^{-3})$ eV allow the small
angle MSW solution to the solar neutrino problem,
\item{ii.)} the  tau neutrino mass, $m_{\nu_\tau} \sim few\ $ eV,
allows the tau neutrino to play the role of the hot dark matter
component  in a mixed dark matter scenario \ref\darkm{
M. Davis, F. J. Summers and D. Schlegel,
Nature 359 (1992) 393; A. N. Taylor and M. Rowan-Robinson, Nature
359 (1992) 396.}, and
\item{iii.)} oscillations between muon and tau netrinos may well be
observable by the collaborations NOMAD and CHORUS at CERN
\ref\dilella{L. DiLella, Nuclear Physics B (Proc. Suppl.) 31 (1993)
319.}. I show the model's predictions compared to present and
future experimental limits in a plot of the
$\sin^2(\Theta_{\nu_\mu\tau})-\Delta(m^2) $ plane.

\noindent The Yukawa matrices of this model are similar to the
well-known Georgi-Jarlskog (GJ) matrices \ref\gj{H. Georgi and C.
Jarlskog, \pl{86}{1979}{297}.} with a few small but important
differences. The family symmetry leads to non-zero entries in the
$\{2,3\}$ and $\{3,2\}$ components of the down quark and charged
lepton Yukawas. These entries have the effect of lowering the
prediction for $|V_{cb}|$ which in the GJ scheme is rather high. The
other difference is a 20\%  correction to
$\lambda_b^{GUT}=\lambda_\tau^{GUT}$ which stems from an
operator that involves $SU(5)$ breaking VEVs. This contribution
lowers the otherwise unacceptably high value obtained for $R =
m_b/m_\tau$ . A more detailed discussion of problematic mass and
angle relations is left to the conclusions.

\newsec{Fields and interactions}

In this section, I present a specific supersymmetric $SO(10)\times
\Delta(75)$ GUT which incorporates the features discussed in the
introduction.
The $\Delta(75)$ family symmetry constrains the allowed Yukawa
couplings of the SM fermions, leading to a modified GJ texture.

\topinsert
\centerline{\epsfbox{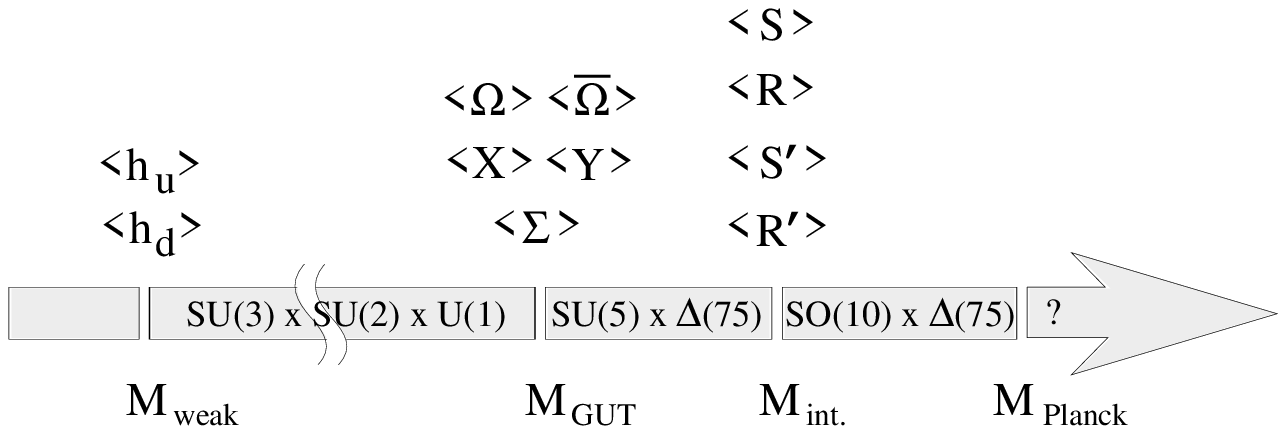}}
\caption{{\bf Fig. 1.} {\it The various mass scales of the model and
the symmetry breaking VEVs.}}
\endinsert

Table 1 lists the superfields involved in the generation of quark and
lepton masses. The three families of the SM are contained in the
superfield $\CF$ .Then there are the fields $\psi$, $ \bar \psi$ and
$\chi$, $ \bar \chi$; they are superheavy and do not acquire VEVs.
Various Higgs fields break the gauge and family symmetries in two
steps. Figure 1 shows the mass scales at which the spontaneous
symmetry breaking takes place. $M_I$ is the scale where
$SO(10)\times U(1)$ is broken to its subgroup $SU(5)$ by the VEVs
of the fields $S,S',R,R'$, and the fields $\Psi, \bar \Psi$ and $\chi,
\bar \chi$ obtain large masses. At $M_{GUT}$ the gauge symmetry is
further broken to the MSSM gauge group by the VEVs of $\Sigma$
and $\Omega$. At the same scale the flavor symmetry is broken by
the VEVs of $X$, $Y$, and $\Sigma$. When the heavy ``matter fields"
$\psi$, $ \bar \psi$ and $\chi$, $ \bar \chi$ are integrated out of the
theory and the flavor symmetry breaking Higgs fields acquire their
VEVs they generate effective Yukawa couplings for the light fields.
These couplings will be suppressed by varying powers of
$$ \epsilon \simeq {<X, Y, \Sigma> \over (M_\psi, M_\chi)} \simeq
{M_{GUT} \over M_I}. $$
Finally, at the weak scale the Higgs doublets acquire their VEVs,
thus giving the masses to the SM quarks and leptons.

\topinsert

\thinsize=.25pt
\thicksize=1.00pt
\begintable
\multispan{1}\tstrut\hfill{Field}\hfill \| \multispan{1}\tstrut\hfill
{$SO(10)$}\hfill | \multispan{1}\tstrut\hfill
{$\Delta\(75\)$}\hfill | \multispan{1}\tstrut\hfill
{$Mass$}\hfill \|\| \multispan{1}\tstrut\hfill{Field}\hfill \|
\multispan{1}\tstrut\hfill
{$SO(10)$}\hfill | \multispan{1}\tstrut\hfill
{$\Delta\(75\)$}\hfill | \multispan{1}\tstrut\hfill
{$Mass$}\hfill \crthick
{$\CF$} \| $16$ | $T_1$ |$ M_W $ \|\|
{$\Sigma$} \| $45$ | $\bar T_4$ |$ M_{GUT} $ \cr
{$\Psi\ ,\bar \Psi$} \| $16\ ,\bar {16}$ | $\bar T_4\ ,T_4$ |$ M_I $ \|\|
{$\Omega\ ,\bar \Omega$} \| $16\ ,\bar {16}$ | $1$ |$ M_{GUT} $ \cr
{$\chi\ ,\bar \chi$} \| $10\ ,10$ | $\bar T_2\ ,T_2$ |$ M_I $ \|\|
{$X$} \| $1$ | $\bar T_3$ |$ M_{GUT} $ \cr
{$S$} \| $45$ | $1$  |$ M_I $ \|\|
{$Y$} \| $1$ | $\bar T_2$ |$ M_{GUT} $ \cr
{$S'$} \| $1$ | $1$ |$ M_I $ \|\|
{$H_u$} \| $10$ | $\bar T_2$ |$ {M_{GUT}}^* $ \cr
{$R$} \| $1$ | $1$ |$ M_I $ \|\|
{$H_d$} \| $10$ | $\bar T_3$ |$ {M_{GUT}}^* $ \cr
{$R'$} \| $1$ | $1$ |$ M_I $ \|\|
{$H_d'$} \| $10$ | $\bar T_3$ |$ {M_{GUT}}^* $ \endtable

\caption{{\bf Table 1.} {\it Fields and representations. Stars point out
that the components of the $H$ fields that correspond to the
electroweak breaking Higgs $h_u$ and $h_d$ stay massless at
$M_{GUT}$.}}

\endinsert

Given the fields and representations in Table 1, the most general
renormalizable superpotential consistent with the symmetries is
\eqn\wpot{\eqalign{
 W_4 = & X \psi   \CF   + \Sigma \bar \psi   \CF   +  H_d \bar\chi \Sigma
+ H_d' \psi \psi \ \cr
&+ \chi\[\CF   \CF   + \CF   \psi \] +H_u \[\CF   \CF   + \CF    \psi   \] ,\cr
}}
where I have suppressed all coupling constants, they are assumed
to be $\CO(1)$.
Several remarks about this superpotential are in order:
\item{1.} I have omitted a $S\bar\chi H_u$ operator; it can be rotated
away by a suitable redefinition of $\chi$ and $H_u$ which carry
identical quantum numbers.
\item{2.} The down type Higgs fields do not have renormalizable
couplings to the SM fermions. This implies that the bottom Yukawa
coupling is automatically suppressed compared to the top coupling,
resulting in low $\tanb=\vev{h_u/h_d}$ and thus avoiding the
problems associated with large $\tanb$\nref\tanbeta{T. Banks,
\np{303}{1988}{172};  A.E. Nelson and L. Randall,
\pl{316}{1993}{516}.} \nref\sarid{R. Rattazzi, U.Sarid, L. Hall, preprint
RU-94-37 (1994) hep-ph-9405313.} \refs{\tanbeta, \sarid}.
\item{3.} The down quark and lepton Yukawa couplings get
contributions from two down type Higgs fields.
Only a linear combination of the third flavor component of $H_d$
and the first flavor component of $H_d'$ remains light after the flavor
symmetry breaking.

\noindent Since the GUT scale and the $SO(10)$ breaking scale are
only a couple of orders of magnitude from the Planck scale, there
are non-negligible contributions to the Yukawa couplings from
operators of dimension greater than four. These operators arise
from gravitational interactions and are suppressed by the
appropriate powers of $M_{P}$:
\eqn\dimfive{
W_{5+6} = {1\over {M_p}}\[\bar\chi Y \bar\Omega \bar\Omega\] +
{1\over {M_p}^2}\[ H_u \CF\CF Y R+  H_d \CF\CF  Y R'\]\ .}
The first term is important for the masses of the right handed
neutrinos, while the dimension six operators contribute to the first
family Yukawa couplings.

In order to generate the MSSM with realistic masses for the quarks
and leptons, it is
necessary to make certain assumptions about the symmetry
breaking pattern. I assume the following:
\item{1.} The fields $S, S', R, R'$ acquire VEVs at the scale $M_I$
which lies somewhere between $M_{GUT}$ and $M_P$, giving large
masses to the $\psi$ and $\chi$ fields. The VEV of $S$ also breaks
$SO(10)$ down to $SU(5)$.
\item{2.} $SU(5)$ is further broken to $SU(3)\times SU(2)\times U(1)$
at $M_{GUT}$
by VEV of $\Sigma$.  Each flavor component of the field $\Sigma$
develops an identical VEV. This also breaks the family symmetry
$\Delta(75)$ to its subgroup $Z_3$.
\item{3.} The family symmetry is broken completely by the fields $X$
and $Y$. $X$ develops a GUT scale VEV along its first component,
thus leaving a $Z_5$ subgoup unbroken, while $Y$ has identical
VEVs in all three components. It is through the VEVs of $X$, $Y$,
and $\Sigma$ that mass mixing between the heavy fermions $\psi$,
$\chi$
and the light fermions $\CF$ is induced.
\item{4.}  The $SU(2)\times U(1)$ breaking VEVs are a little more
complicated. I assume that only the $Y=-1/2$ weak doublet from
$(H_u)_3$ and the weak $Y=+1/2$ doublet contained in a linear
combination of  $(H_d)_3$ and $(H_d')_1$
remain lighter than $M_{GUT}$ and participate in electroweak
symmetry breaking.
In the following, I denote the light Higgs doublets by $h_u$ and
$h_d$.

\topinsert
\centerline{\epsfbox{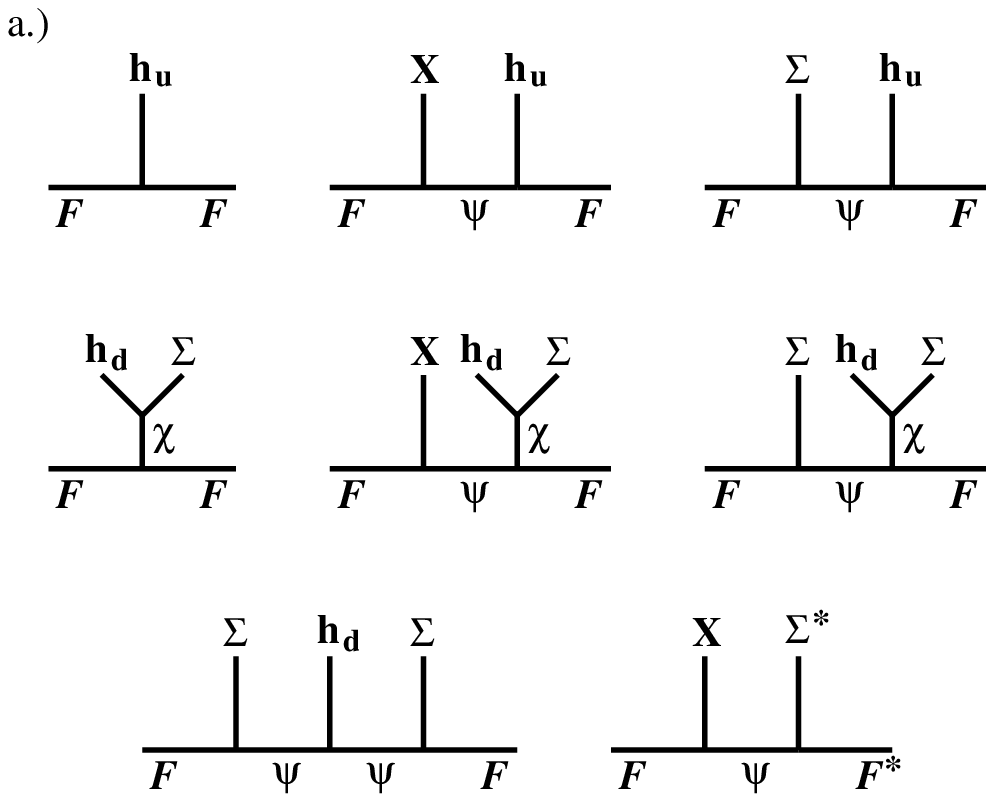}}
\centerline{\epsfbox{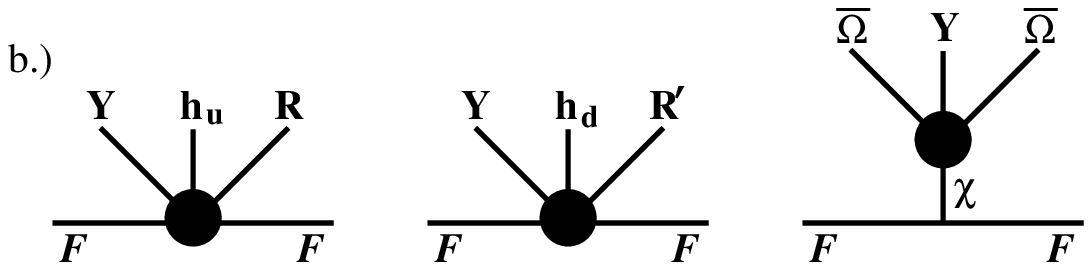}}
\caption{{\bf Fig. 2.} {\it Leading  supergraph contributions to quark
and
lepton Yukawa couplings.  Internal lines indicate $\psi$, $\bar \psi$,
$\chi$ and $\bar \chi$ superfields. External lines are the
light fermions $\CF$ and various Higgs fields. Black blobs represent
non-renormalizable interactions suppressed by the appropriate
powers of $M_P$. The last diagram in fig. 2.a. corresponds to an
effective D-term.}}
\endinsert

\noindent One can now determine the resulting effective Yukawa
couplings just below $M_{GUT}$
by calculating the diagrams with renormalizable couplings (Figure
2.a.) and with non-renormalizable couplings (Figure 2.b.).

\newsec{Quark and charged lepton Yukawa couplings}

One sees from the diagrams in Figure 2.a. that only the top quark
field has a renormalizable coupling to the weak scale Higgs fields. All
other quark and lepton Yukawa couplings involve flavor symmetry
breaking and are suppressed. The second diagram contributes to
the $\{2,3\}$ and $\{3,2\}$ entries of the up Yukawa matrix, and is
suppressed by a factor of $\epsilon_x = {<X> \over M_\psi}  \sim
{M_{GUT} \over M_I}$. The third diagram does not contribute to
$Y_u$ because of a vanishing $SO(10)$ Clebsch Gordan coefficient.
The fourth through sixth diagrams are the corresponding diagrams
for the down quark and charged lepton sector. They are suppressed
by $\epsilon_\sigma = {<\Sigma> \over M_\chi} \sim {M_{GUT} \over
M_I}$ compared to the first three diagrams.
The seventh diagram gives an additional contribution to the $\{3,3\}$
entries of the down and charged lepton Yukawa couplings. This
contribution is not $SU(5)$ symmetric and splits $m_b$ and
$m_\tau$ by an amount of $\CO(20\%)$. The eighth diagram
corresponds to a flavor off-diagonal effective D-term and requires
wavefunction renormalization. However, wave function
renormalization is negligible in this model.

The first and second diagrams in Figure 2.b. contribute to the
$\{1,2\}$ and $\{2,1\}$ entries of the up and down Yukawa coupling
matrices\foot{The third diagram does not contribute to SM fermion
masses. It generates large masses for the right handed neutrinos.}.
They are suppresssed by factors of $\delta = {<R><Y> \over M_P^2}
\sim {M_{GUT} \over M_P} {M_I \over M_P}$.

Taking into account $SO(10)$ Clebsch Gordan coefficients, one
obtains the following Yukawa coupling matrices
\eqn\yredef{\eqalign{
&Y_u = \(\matrix{
{0} &{\delta_u} & {0}\cr
{\delta_u}& {0}& {\epsilon_x}\cr
{0}& {\epsilon_x}& {A}\cr}\) \ ,\qquad
Y_d = \(\matrix{
{0} &{\delta_d  e^{i\phi_1}} & {0}\cr
{\delta_d  e^{-i\phi_1}}& {\frac43 \epsilon_\sigma \epsilon_{\sigma'}
e^{i\phi_2}}& {\frac13 B \epsilon_x\epsilon_\sigma e^{i\phi_3}}\cr
{0}& { B \epsilon_x\epsilon_\sigma e^{-i\phi_3}}&
{|\epsilon_\sigma+\frac13 C {\epsilon_{\sigma'}}^2|}\cr}\) \ ,\qquad \cr
&Y_l =  \(\matrix{
{0} &{\delta_d e^{i\phi_1}} & {0}\cr
{\delta_d e^{-i\phi_1}}& {-4 \epsilon_\sigma \epsilon_{\sigma'}
e^{i\phi_2}}& {-B \epsilon_x\epsilon_\sigma e^{i\phi_3}}\cr
{0}& {-\frac13 B \epsilon_x\epsilon_\sigma  e^{-i\phi_3}}&
{|\epsilon_\sigma+3C { \epsilon_{\sigma'}}^2|}\cr}\) \ ,}}

\noindent where $Y_u^{ij}$, $Y_d^{ij}$, and $Y_l^{ij}$ are the
coefficients of the effective operators
$h_uQ_iU_j^c $,  $h_dQ_iD_j^c $, and $h_dL_iE_j^c $ respectively,
and where I have defined
\eqn\defxsi{\eqalign{\epsilon_x &= {\vev{X_1}\over M_\psi}\ ,\qquad
\epsilon_\sigma = {\vev{\Sigma_1}\over M_\chi} \ ,\qquad
\epsilon_{\sigma'} = {\vev{\Sigma_1}\over M_\psi}=
{\vev{\Sigma_2}\over M_\psi} \ ,\cr
\delta_d &= {\vev{Y_3}\vev{R'}\over M_{P}^2}\ ,\qquad \delta_u =
 {\vev{Y_2}\vev{R}\over M_{P}^2}\ .}}
All the parameters denoted with capital letters are $\CO(1)$. The
parameters $\epsilon_x, \epsilon_\sigma$, and $\epsilon_{\sigma'}$
are expected to be $\CO(10^{-1})$ or $\CO(10^{-2})$ from their
definitions. The $\delta$'s are $\CO(10^{-3})$ or $\CO(10^{-4})$.
Unphysical phases have been rotated away, and I have neglected
the small difference in phase between the $\{3,3\}$ entries of $Y_d$
and $Y_l$. The remaining phases are expected to be $\CO(1)$.
I have given only the leading
contributions to each entry, and ignore the negligible contributions
from
wavefunction renormalization to the $\{13\}$, $\{31\}$ and $\{11\}$
entries. One sees that there is a natural
hierarchical structure to the masses, and that down-type quarks
and leptons are
automatically a factor of $\epsilon_\sigma$ more weakly coupled to
the Higgs doublet than are up-type quarks, thus predicting small
$\tanb$. Notice the $SO(10)$ Clebsch factors \anderson\  appearing
in the matrices:
\item{1.} Factors of $\frac13$ in the $\{2,3\}$ and $\{3,2\}$ entries of
$Y_D$ and $Y_L$.
\item{2.} The third diagram in Fig. 2.a. does not contribute to $Y_u$
because of a zero $SO(10)$ Clebsch factor while the corresponding
diagram for the down sector gives a factor of 3 difference between
the magnitudes of the $\{22\}$ entries in $Y_d$ and $Y_l$. The factor
of -3 plays the same role as the factor of -3 in the GJ mass matrices.
\item{3.} The corrections to the $b$ and $\tau$ Yukawa couplings
from the seventh diagram have different Clebsch factors, thus
splitting $\lambda_b$ and $\lambda_\tau$ at $M_{GUT}$.
\item{4.} The gravitationally induced interactions which contribute to
the  $u$, $d$, and $e$ masses as well as to the Cabbibo angle do not
contain any $SO(10)$ Clebsch factors.

\newsec{Renormalization group evolution and numerical
predictions}

I now determine the parameters of the Yukawa coupling matrices in
\yredef\  by running them to the scale of the respective fermion
masses, diagonalizing the mass matrices and matching onto the
measured masses and mixing angles. For the evolution between
$M_{GUT}$ and $m_t$ I use one-loop MSSM renormalization group
equations. Between $M_{GUT}$ and the scale of the Majorana
masses of the right handed neutrinos $M_N$, which I take at
$10^{12}$ GeV, one also needs to include the contributions to the
running from the neutrino Yukawa coupling
$\lambda_{\nu_\tau}$\foot{For simplicity, I assume $M_{SUSY} =
m_t$. I also ignore contributions from $\lambda_b$ and
$\lambda_\tau$ to the evolution equations, they are negligible in a
small $\tanb$ scenario. I have checked that using two loop
renormalization group equations does not change the results
significantly.}. Below the scale of the top quark mass I utilize
three-loop QCD and one-loop QED scaling factors which I adopt
from Babu and Mohapatra \ref\babuetas{K. Babu and R. Mohapatra,
preprint BA-94-56 (1994) hep-ph-9410326.}:
$(\eta_{u},\eta_{d,s},\eta_{c},\eta_{b},\eta_{e,\mu},\eta_{\tau})=(2.49,
.48,2.17,1.55,1.02,1.02)$, where $\eta_f = m_f\(m_f\) / m_f\(m_t\)$ for $f
= c$, $b$, $\tau$, and $\eta_f = m_f\(1\ GeV\) / m_f\(m_t\)$ for the light
fermions.
The experimental input parameters are listed in Table 2.

\topinsert

\thinsize=.25pt
\thicksize=1.00pt
\begintable
\multispan{1}\tstrut\hfill{$m_e$}\hfill \| \multispan{1}\tstrut\hfill
{$m_\mu$}\hfill | \multispan{1}\tstrut\hfill
{$m_\tau$}\hfill | \multispan{1}\tstrut\hfill
{$m_c$}\hfill | \multispan{1}\tstrut\hfill
{$m_b$}\hfill | \multispan{1}\tstrut\hfill
{$m_t$}\hfill | \multispan{1}\tstrut\hfill
{${\vert V_{ub}\vert  \over \vert V_{cb}\vert }$}\hfill |
\multispan{1}\tstrut\hfill
{$\vert V_{cb}\vert $}\hfill | \multispan{1}\tstrut\hfill
{$\vert V_{us}\vert $}\hfill \crthick
$5.11\ 10^{-4}$ \| 0.106  | 1.78 | $1.3 \pm 0.3$ | $4.3 \pm 0.2$ | $174
\pm 16$ | $0.08 \pm 0.02$ | $0.040 \pm 0.005$ | 0.221 \endtable
\caption{{\bf Table 1.} {\it Experimental input parameters are taken
from \ref\PDG{Particle Data Group, \physrev{D50}{1994}{1173}, and
references therein.}. Quark masses are displayed in units of GeV.}}

\endinsert

The renormalization procedure for the Yukawa couplings is well-known
and has been performed in a number of publications \nref\runners{V.
Barger, M. S. Berger and P. Ohmann,
\physrev{D47}{1993}{1093}; \physrev{D47}{1993}{2038}; K. S. Babu,
Z. Phys. C 35 (1987) 69; K. S. Babu, C. N. Leung, J. Pantaleone,
\pl{319}{1993}{191}.} \refs{\attempts - \famsym, \runners}.
I only mention some important features.
Most models based on $SU(5)$ or $SO(10)$ lead to the boundary
condition $\lambda_b^{GUT}=\lambda_\tau^{GUT}$, and one
determines $\lambda_t^{GUT}$ through its important contribution to
the running of $R(\mu)=m_b(\mu)/m_\tau(\mu)$. Using a
representative value for $\alpha_s(M_Z)=0.12$, one finds a very high
value for $\lambda_t^{GUT} \sim 3$. While this possibility cannot be
ruled out, it does constitute a serious problem to any predictive GUT
extension because the large top Yukawa coupling becomes infinite
closely above the GUT scale. For example, $\lambda_t^{GUT} = 3$
leads to a Landau pole at 2 $M_{GUT}$ in both  $SU(5)$ and
$SO(10)$. As a result, one loses predictivity completely because
now one expects higher dimension operators ``suppressed" by
factors of $M_{GUT}/2M_{GUT}$ to play an important role.
Demanding perturbativity up to $M_{Planck}$ requires
$\lambda_t^{GUT} \le 1.3$ in this model, and one is forced to give up
and correct the $SU(5)$ relation
$\lambda_b^{GUT}=\lambda_\tau^{GUT}$.
When including the partially cancelling contributions to the running
from both $\lambda_{\nu_\tau}^{GUT} =1.3$ and $\lambda_t^{GUT} =
1.3$
I find $R(M_{GUT})=0.85$ from the experimental input \foot{ The
connection between the mass scale of the right handed neutrinos
and the $m_b/m_\tau$ ratio has been pointed out in \ref\hitoshi{A.
Brignole, H. Murayama, R. Rattazzi, preprint LBL-35774 (1994)
hep-ph-9406397.}.}.
In the following, I fix $\lambda_t^{GUT} = 1.3$ in order to maximize its
contribution to the running of $R$. The predictions of the model are
not very sensitive to variations of $\lambda_t$ as long as
$\lambda_t^{GUT} \le 1.3$.

I now extract $\tanb, \epsilon_x, \epsilon_\sigma, \epsilon_{\sigma'}$
from $m_t, m_\tau, m_\mu, m_c$, respectively.  The parameter $B$
can be determined from $|V_{cb}^{GUT}| = {\epsilon_x \over  A}|
1-{A B \over  3} e^{i \phi_3}|  $. This constrains $0.62\le|B|\le4.2$, but
I will limit $B \le 2$ because i.) larger values are disfavored by the
experimental limits on $\nu_\mu \rightarrow \nu_\tau$ oscillations as
I will show in the following section, and ii.) a value of $B \sim 1$ is
favored from a theoretical viewpoint since $B$ is defined as a
combination of  $\CO(1)$ coupling constants.
 From the equations for $|V_{cb}|$ and $|V_{us}|$ one can also
determine the phases $|\phi_3|$ and $|\phi_1-\phi_2|$. However, this
does not suffice to predict CP violation because of the
unconstrained phase $|\phi_1+\phi_2|$. Finally,  $\delta_u$ and
$\delta_d$ are determined from $| V_{ub}|  / |V_{cb}| $ and $m_e$.
Numerically these parameters are

\eqn\results{\eqalign{
A&=1.3, \quad 0.62 \leq B \leq 2,
\quad |1+3C{{\epsilon_{\sigma'}}^2 \over \epsilon_\sigma}|=1.20, \cr
\epsilon_x&=5.6\ \pm 0.8 \ 10^{-2}, \quad \epsilon_\sigma=1.30 \
10^{-2}, \quad \epsilon_{\sigma'}=1.67 \ 10^{-2}, \cr \delta_u &= 1.9 \
10^{-4}, \quad \delta_d = 0.51 \ 10^{-4}, \quad \tanb=1.94\ . \cr }}
\noindent The parameters A, B, and C are of $\CO(1)$, as expected.
This means that $\Delta(75)$ is ``working properly", that is, no
unnaturally large or small couplings in the superpotential are
necessary. The hierarchy of Yukawa couplings is entirely explained
as powers of $\Delta(75)$ symmetry breaking VEVs over intermediate
particle masses or the Planck scale.

The model leads to three predictions in the quark and charged
lepton sector:
\eqn\smu{
m_s ={\left| 1-2\xi\right|  \over 3 \eta}{\eta_s \over \eta_\mu}m_\mu
188 \pm 3 \% B^2\ {\rm MeV} \ ,}
\eqn\preds{
{m_d \over m_s}=9 \left| 1+2\xi\right| ^2 {m_e \over m_\mu} = {1 \over
22.9} \pm 6 \% B^2 \ ,  }
\eqn\mup{
m_u=\left({V_{ub}  \over V_{cb}}\right)^2{\eta_u \over \eta_c} m_c =
9.5 \pm 5.2\ {\rm MeV}\ . }
Here $\eta=0.45$ is an evolution factor that accounts for the running
of ${ m_s \over m_\mu}$ from the GUT scale down to the weak scale.
$\eta_u$, $\eta_c$, $\eta_\mu$, and $\eta_s$ have been given before,
and $\xi={B^2 {\epsilon_x}^2 \over 12
\epsilon_{\sigma'}}e^{-i\gamma}$ is small. It varies between $6.0\
10^{-3}$ and  $6.2\  10^{-2}$ as $B$ is varied from $0.62$ to $2$. The
predictions should be compared to the estimates from chiral
perturbation theory \PDG. The value for $m_d/m_s$ agrees very well,
while the value for $m_u/m_d=1.16 \pm 55\%$ is quite large and is
only consistent because of the large uncertainties in the prediction
which stem from the experimental error bars of the input value  for $|
{V_{ub}  \over V_{cb}}|$.

\newsec{Neutrino masses}

The field $\CF$ that contains all the SM fermions
also contains an $SU(3) \times SU(2) \times U(1) $ singlet field that
plays the role of the Dirac partner $N^c$ of the left handed neutrino
in the lepton doublet. The $SO(10)$ symmetry relates the neutrino
Yukawa coupling matrix $Y_\nu^{ij}$ to the up quark Yukawa matrix
by known Clebsch Gordan coefficients

\eqn\ynueffi{
Y_\nu = \(\matrix{
{0} &{\delta_u} & {0}\cr
{\delta_u}& {{8 \over 5} {\epsilon_{\sigma'} \over B }  e^{i(\phi_2
+\phi_3)}}& {\frac15 \epsilon_x}\cr
{0}& {\frac13 \epsilon_x}& {A}\cr}\) \ .}

All the components are given in terms of parameters already
determined from the  quark and charged lepton sector. Note that
unlike the corresponding up Yukawa matrix, the $\{2,2\}$ component
of $Y_\nu^{ij}$ does not have a vanishing Clebsch factor. This leads
to an interesting modification of the usual quadratic seesaw
mechanism \nref\neutex{H. C. Cheng, M. S. Gill, L. J. Hall,
\physrev{D49}{1994}{4826}; K. S. Babu and Q. Shafi,
\pl{311}{1993}{172}.}
\refs{\seesaw,\neutex}.
The effective Majorana mass of the neutrinos as we would measure it
in a successful neutrino oscillation experiment is then given by

\eqn\majmass{
M_{\nu}=-\[{\upsilon \sin \beta \over 2}\]  Y_\nu \, M_{N}^{-1}  \,
{Y_\nu}^T\ }
where $M_N^{-1}$ is the inverse of the Majorana mass matrix of the
heavy right handed neutrinos.
In general $M_N$ is completely arbitrary. But in a model such as this
one where the fermions transform as irreducible triplet
representations of the family symmetry, the form of $M_N$ is very
restricted. $\Delta(75)$ predicts it to be either proprotional to the
unit matrix or else completely off-diagonal with all identical entries.
In either case the resulting $M_N^{-1}$ is non-hierarchical and
completely determined except for the overall mutiplicative mass
scale.
In this model, the third diagram in Figure 2.b. leads to $M_N$
proportional to the unit matrix with an overall factor $\vev{Y}
\vev{\Omega}^2/M_PM_I \sim M_{GUT}^3/M_PM_I \sim 10^{12}$
GeV.  Diagonalizing, I find the following predictions for the neutrino
mass ratios and lepton mixing angles:
\eqn\neupred{\eqalign{
{m_{\nu_\mu} \over m_{\nu_\tau}}&= \left({8 \over 5}
{\epsilon_{\sigma'} \eta_\nu \over A B} \right)^2 \simeq 6.5\ 10^{-4}
B^{-2}\ ,\cr
{m_{\nu_e} \over m_{\nu_\mu}}&= {\delta_u^4 \eta_\nu^4 \over A^4}
\left({m_{\nu_\tau}  \over m_{\nu_\mu}}\right)^2 \simeq 2.6\
10^{-9}B^4\ ,\cr   }}
\eqn\lepang{\eqalign{
|\Theta_{\nu_e\mu}| &= \sqrt{m_e \over m_\mu} \simeq 0.069 \pm
0.007 B , \cr
|\Theta_{\nu_\mu\tau}| &= B \epsilon_x \left| \eta_N + {\eta_\nu e^{-i
\beta} \over 5 A B} \right| \simeq B \epsilon_x \left( \eta_N + {\eta_\nu
\over 5 A B} \right)  \simeq 0.059 B + 0.011, \cr
\left|{\Theta_{\nu_e\tau} \over \Theta_{\nu_\mu\tau}}\right| &=
\left({m_{\nu_e} \over m_{\nu_\mu}}\right)^{1/4} \simeq 0.007 B ,\cr}}

\topinsert
\centerline{\epsfbox{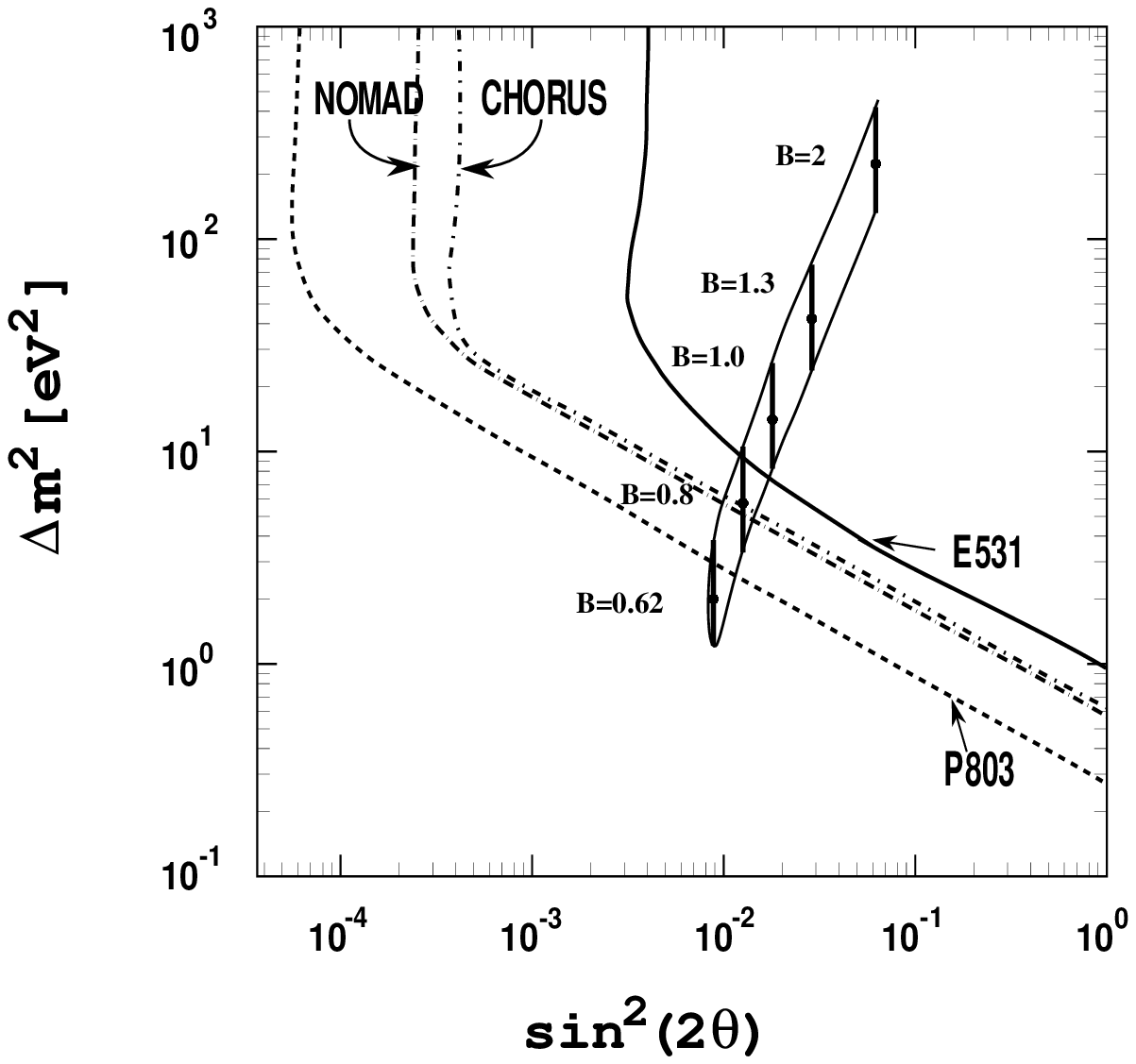}}
\caption{{\bf Fig. 3.} {\it Predictions for  $m_{\nu_\tau}^2$ and
$\sin^2(2\Theta_{\nu_\mu\tau})$  compared to limits from past and
future $\nu_\mu - \nu_\tau$ oscillation experiments \dilella. The
predictions for $\Theta_{\nu_\mu\tau}$ and $m_{\nu_\tau}$ increase
with increasing $B$. }}
\endinsert

\noindent with the evolution factors $\eta_\nu = 1.24$ and $\eta_N =
1.06$. The overall mass scale is approximately given by
$m_{\nu_\tau}\sim few $ eV and therefore $m_{\nu_\mu} \sim
10^{-3}$eV . While the prediction for $m_{\nu_\tau}$ allows the tau
neutrino to play the role of a hot dark matter candidate in a mixed
dark matter scenario, the model also offers a solution to the solar
neutrino problem via $\nu_e \leftrightarrow \nu_\mu$ oscillations.

This suggests that we use the experimental value for $\Delta m^2$
from the MSW solution to the solar neutrino problem,
$m_{\nu_\mu}=1.8\ 10^{-3} - 3.5\ 10^{-3}$ eV, as input in order to fix
the overall masscale \ref\msw{N. Hata and P. Langacker, preprint
UPR-0625-T (1994) hep-ph-9409372.}. The resulting predictions for
$\nu_\mu \rightarrow \nu_\tau$ oscillations are plotted in Figure 3.
together with the present limits from accelerator oscillation
experiments and the expected sensitivities for the new generation of
experiments, NOMAD and CHORUS at CERN. One finds that large
values of $B \geq 1.2$ are already ruled out and the exciting
prospect that NOMAD and CHORUS may soon see the first direct
evidence for neutrino oscillations.

For completeness, I also mention that the model's predictions for
$\nu_e \leftrightarrow \nu_\tau$ oscillations are far from current
experimental limits due to the small $\Theta_{\nu_e\tau}$ mixing
angle, and that they are consistent with more stringent limits derived
from heavy element nucleosynthesis in supernovae \ref\fuller{Y. Z.
Qian {\it et al.}, \prl{71}{1993}{1965}; Y. Z. Qian and G. Fuller,
preprint DOE-ER-40561-150 (1994) astro-ph-9406073.}.

\newsec{Conclusions}

In this letter I have presented a new SUSY-GUT model which
predicts fermion masses and mixing angles. The non-Abelian family
symmetry group of the model explains the observed hierarchical and
diverse spectrum of masses and angles in terms of the hierarchy
between the mass scales $M_{GUT}$, $M_{Planck}$, and an
intermediate scale $M_I$ where $SO(10)$ is broken down to $SU(5)$.
The numerical predictions arise because the $SO(10)$ symmetry
relates entries from different Yukawa matrices via Clebsch Gordan
coefficients.

Particularily interesting is the $SO(10)$ Clebsch structure in the
 $\{2,2\}$,  $\{2,3\}$, and $\{3,2\}$ components of the Yukawa
matrices. These Clebsches, while ensuring consistency with
measured SM masses and angles, also show up in the neutrino
sector and lead to predictions which are more successful than the
naive quadratic seesaw relations.

However, while the model meets the goal of generating a Yukawa
texture that is predictive and in accord with all experimental data, it
does so only at the cost of a rather complicated symmetry breaking
sector. This is due to problems that seem to be generic. Many of the
mass and angle relations that have been derived from various
``standard textures" such as the Fritzsch texture or the GJ texture
do not work at the level of precision at which we know SM
parameters
today. I list three such problematic relations:

\item{1.} The most ``annoying" problem in the context of an $SU(5)$
or $SO(10)$ theory is that $\lambda_b^{GUT}=\lambda_\tau^{GUT}$
unification does not lead to a believable prediction for
$m_b/m_\tau$. The problem is the following. The renormalization
group equation for  $R(\mu) =  m_b(\mu)/m_\tau(\mu)$ depends
crucially on a large top Yukawa coupling\foot{I am implicitely
assuming small $\tanb$ in ignoring contributions from $\lambda_b$
and $\lambda_\tau$. For large $\tanb$ the prediction of
$m_b/m_\tau$ has recently been shown to be problematic as well
\sarid.}. A prediction consistent with experiment requires
$\lambda_t^{GUT} \sim 3$. However, such a large Yukawa coupling
leads to a Yukawa Landau pole closely above the GUT scale ($2
M_{GUT}$). This opens a Pandora's box of nearly unsuppressed
higher dimensional operators which are expected to arise from the
non-perturbative physics, and predicitvity is lost completely.
Alternatively, one could limit $\lambda_t^{GUT} \le 1.3$ and avoid a
Landau pole below $M_{Planck}$ at the cost of giving up
 $\lambda_b^{GUT}=\lambda_\tau^{GUT}$. However, the necessary
$\CO(15\%)$ corrections to $R$ introduce a new parameter and loss
of predictivity. Also, this fix renders the model more complicated
because it is not easy to move the ``cornerstone" of an $SU(5)$
Yukawa theory which really sits at $R(M_{GUT})=1
$\foot{$\lambda_b^{GUT}$ and $\lambda_\tau^{GUT}$ can only be
split by the VEV of a $\bar {45}$ of $SU(5)$. The $\bar {45}$ could
either be an additional down type Higgs field (dangerously large
contribution to the gauge $\beta$ function), or it could be a more
complicated product of Higgs fields.}. In an $SO(10)$ theory the
situation is further  worsened by a cancellation of the top
contribution to the $\beta$ function for $R$ by an identical
contribution from $\lambda_{\nu_\tau}$ which enters with opposite
sign. The lower the scale of the right handed neutrino masses, the
larger (worse) the prediction for $R$. For a review see \hitoshi.

\item{2.} A problem for models based on the GJ texture is the high
value predicted for $V_{cb} \simeq 0.050$ \foot{This prediction is
especially high in the case of Yukawa trinification (large $\tanb$)
\refs{\largetanb, \babuetas}.}. In the context of family symmetries this
relation finds an easy and rather
natural fix via additonal entries in the down quark matrix.

\item{3.} The last problem I want to mention is the relation
$\sqrt{\frac{m_u\eta_c}{m_c\eta_u}}=\left|\frac{V_{ub}}
{V_{cb}}\right|$.
It arises from all textures with zeros in the $\{1,1\}$, $\{1,3\}$, and
$\{3,1\}$ components of the Yukawa matrices. This relation, while not
being excluded, is disfavored because it predicts a rather high value
for $m_u \simeq 9.5 \pm 5.2$ MeV. If combined with the GJ prediction
for $m_d \sim 8$ MeV this results in $m_u/m_d =1.2 \pm 0.6$.

\noindent It is encouraging to see that non-Abelian family
symmetries lead to interesting textures with predictions that are very
similar to the real world. However, it is frustrating to see that as the
SM parameters are measured more and more accurately, the models
that are in agreement with all data become increasingly complicated
and less appealing.  A successful predictive SUSY $SO(10)$ or
$SU(5)$ GUT model will have to include a solution to the
$m_b/m_\tau$ problem and probably new Yukawa matrix textures.

\vskip .4in
\centerline{{\bf Acknowledgements}}

I wish to express my gratitude to David Kaplan for numerous
valuable discussions and encouragement. This work was supported
by the Department of Energy
under Grant No. DE-FG06-90ER40561.

\listrefs
\vfill\eject

\bye